\documentclass[pra,aps,showpacs,twocolumn]{revtex4}
\usepackage{amssymb}
\usepackage{amsmath}
\usepackage{mathtools}
\usepackage[dvipdfm]{hyperref}
\usepackage{hyperref}

\allowdisplaybreaks
\begin{document}

\title{Quantifying coherence with respect to general measurements
via quantum Fisher information}
\author{Jianwei Xu}
\email{xxujianwei@nwafu.edu.cn}
\affiliation{College of Science, Northwest A\&F University, Yangling, Shaanxi 712100,
China}


\begin{abstract}
The standard quantum coherence theory is defined with respect to an
orthonormal basis of a Hilbert space. Recently, Bischof, Kampermann and Bru%
\ss\ generalized the notion of coherence into the case of general
measurements, and also, they established a rigorous resource theory of
coherence with respect to general measurements. In this paper, we propose
such a coherence measure with respect to general measurements via quantum
Fisher information, and provide an application of this measure.
\end{abstract}

\maketitle

\section{Introduction}
Coherence is a fundamental ingredient in quantum physics. Since Baumgratz,
Cramer and Plenio \cite{BCP-2014-PRL} established a rigorous framework for
quantifying coherence (we call this framework BCP framework), fruitful
results about coherence have been achieved both in theories and experiments,
for reviews see \cite{Plenio-2016-RMP}\cite{Fan-2018-PhysicsReports}. For a
quantum system associating with a $d$-dimensional Hilbert space $H,$ the BCP
framework considers the coherence defined with respect to an orthonormal
basis $\{|j\rangle \}_{j=1}^{d}$, we call such coherence standard
coherence. An orthonormal basis $\{|j\rangle \}_{j=1}^{d}$ corresponds to a
rank-$1$ projective measurement $\{|j\rangle \langle j|\}_{j=1}^{d},$ hence
we may ask whether or not we can extend the standard coherence to the case of general
measurements. A general measurement, or called a POVM (Positive
Operator-Valued Measurement) \cite{Nielsen-2000-book}, is described by a set
of positive-definite operators $E=\{E_{j}\}_{j=1}^{n}$ satisfying $%
\sum_{j=1}^{n}E_{j}=I_{d}$ with $I_{d}$ the identity operator on $H.$
Recently, Bischof, Kampermann and Bru\ss\ generalized the notion of coherence
into the case of general measurements \cite{Brub-2019-PRL}, and established
a framework of quantifying coherence with respect to general measurements
(we call this framework BKB framework) \cite{Brub-2021-PRA}. BKB framework
includes BCP framework as a special case when the POVM $E$ is a rank-$1$
projective measurement. We call the coherence under BKB framework as POVM
coherence.

Many standard coherence measures have been found and some of them possess
physical interpretations and applications \cite{Plenio-2016-RMP}\cite%
{Fan-2018-PhysicsReports}. So we expect that some of these standard
coherence measures can be generalized to be the corresponding POVM coherence
measures. In fact, several standard coherence measures have been properly
generalized to be POVM coherence measures, such as the relative entropy of
POVM coherence \cite{Brub-2019-PRL,BCP-2014-PRL,Yang-2016-PRL}, robustness
of POVM coherence \cite{Brub-2021-PRA,Adesso-2016-PRL,Adesso-PRA-2016}, $%
l_{1}$ norm of POVM coherence \cite{Brub-2021-PRA,BCP-2014-PRL,Xu-2020-PRA},
and the POVM coherence based on the Tsallis entropy \cite{Xu-2020-PRA,Rastegin-2016-PRA,Yu-2017-PRA,Yu-2018-SR,Xiong-2018-PRA}. Also, there is a
physical interpretation for the relative entropy of POVM coherence \cite{Brub-2021-PRA}. Recently, the relationship between POVM coherence and entanglement was investigated \cite{Wu-2021-PRA}.

In this work, we propose a POVM coherence measure via the quantum Fisher
information (QFI) and provide an application of it. Quantum Fisher
information is a core concept in quantum metrology \cite{Liu-2019-JPA}, we
hope the POVM coherence measure via the quantum Fisher information proposed
in this work has potential applications in quantum metrology. This paper is
organized as follows. In section II, we review the framework of block
coherence and the BKB framework for POVM coherence. In section III, we
propose the POVM coherence measure via QFI and prove that this is a valid
POVM coherence measure. In section IV, we provide a physical application for
the POVM coherence measure via QFI. Section V is a brief summary.

\section{Block coherence and POVM coherence}

In this section we review the standard coherence, block coherence and POVM
coherence, they can be viewed as special cases of quantum resource theory
\cite{Gour-2019-RMP}. We first review the BCP framework about standard
coherence. BCP framework is defined with respect to a fixed orthonormal
basis $\{|j\rangle \}_{j=1}^{d}$ of Hilbert space $H.$ In BCP framework, a
state $\rho $ is defined as an incoherent state if
\begin{eqnarray}
\rho =\sum_{j=1}^{d}p_{j}|j\rangle \langle j|,  \label{eq2.1}
\end{eqnarray}%
where $\{p_{j}\}_{j=1}^{d}$ is a probability distribution. Equivalently, an
incoherent state is a diagonal state in the fixed orthonormal basis $%
\{|j\rangle \}_{j=1}^{d},$ or we say, a state $\rho $ is incoherent if and
only if
\begin{eqnarray}
\langle j|\rho |k\rangle =0,j\neq k.  \label{eq2.2}
\end{eqnarray}%
In BCP framework, a channel \cite{Nielsen-2000-book} $\phi $ is called
incoherent if $\phi $ allows a Kraus operator decomposition $\phi
=\{K_{l}\}_{l}$ such that $K_{l}\rho K_{l}^{\dagger }$ is diagonal in the
fixed orthonormal basis $\{|j\rangle \}_{j=1}^{d}$ for any $l$ and any
incoherent state $\rho .$ We call such Kraus operator decomposition $\phi
=\{K_{l}\}_{l}$ an incoherent decomposition.

Block coherence is a generalization of standard coherence to the case of
projective measurements \cite{Aberg-2006-arxiv,Brub-2021-PRA}. For a
projective measurement $P=\{P_{j}\}_{j},$ that is, $P_{j}P_{k}=\delta
_{jk}P_{j}$, $\sum_{j}P_{j}=I_{d},$ a state $\rho $ is called block
incoherent (with respect to $P$) if and only if
\begin{eqnarray}
P_{j}\rho P_{k}=0,\forall j\neq k.   \label{eq2.3}
\end{eqnarray}%
A channel $\phi $ is called block incoherent if $\phi $ allows a Kraus
operator decomposition $\phi =\{K_{l}\}_{l}$ such that for any block incoherent state $\rho,$
\begin{eqnarray}
P_{j}K_{l}\rho K_{l}^{\dagger }P_{k}=0,\forall l,\forall j\neq k.   \label{eq2.4}
\end{eqnarray}%
We call such decomposition $\phi =\{K_{l}\}_{l}$ an block incoherent
decomposition.

In Ref. \cite{Brub-2021-PRA}, the authors established a rigorous framework
for quantifying the block coherence that a real-valued functional $C(\rho
,P) $ with respect to $P$ is a block coherence measure if it satisfies (B1)
to (B4) below.

(B1). Nonnegativity: $C(\rho ,P)\geq 0,$ and $C(\rho ,P)=0$ if and only if $%
\rho $ is block incoherent.

(B2). Monotonicity: $C(\phi (\rho ),P)\leq C(\rho ,P)$ for any state $\rho $
and any block incoherent channel $\phi .$

(B3). Strong monotonicity: $\sum_{l}$tr$(K_{l}\rho K_{l}^{\dagger })C(\frac{%
K_{l}\rho K_{l}^{\dagger }}{\text{tr}(K_{l}\rho K_{l}^{\dagger })},P)\leq
C(\rho ,P)$ for any state $\rho $ and any block incoherent channel $\phi $
with $\phi =\{K_{l}\}_{l}$ a block incoherent decomposition.

(B4). Convexity: $C(\sum_{l}p_{l}\rho _{l},P)\leq \sum_{l}p_{l}C(\rho
_{l},P) $ for any states $\{\rho _{l}\}_{l}$ and any probability
distribution $\{p_{l}\}_{l}.$

One can check that (B3) and (B4) together imply (B2), and also (B1) to (B4)
returns to the case of standard coherence when $P=\{P_{j}\}_{j}$ is rank-$1$
\cite{BCP-2014-PRL}, i.e., rank$P_{j}=1$ for all $j.$

In Ref. \cite{Xu-2020-PRA}, the authors proposed (B5) below.

(B5). Block additivity:
\begin{eqnarray}
C(p\rho _{1}\oplus (1-p)\rho _{2},P)=pC(\rho
_{1},P)+(1-p)C(\rho _{2},P),   \label{eq2.5}
\end{eqnarray}
here $p\in \lbrack 0,1],$ states $\rho _{1}$ and $\rho _{2}$ satisfy that
there exists a partition $P=\{P_{l}\}_{l}=\{P_{l_{1}}\}_{l_{1}}\cup
\{P_{l_{2}}\}_{l_{2}}$ such that $\{P_{l_{1}}\}_{l_{1}}\cap
\{P_{l_{2}}\}_{l_{2}}=\varnothing ,$ $\rho _{1}P_{l_{2}}=\rho
_{2}P_{l_{1}}=0 $ for any $l_{1}$ and $l_{2}.$

It has been proved that (B2)+(B5) is equivalent to (B3)+(B4) \cite%
{Xu-2020-PRA} hence (B5) provide an alternative way for verifying a block
coherence measure. For many cases verifying (B2)+(B5) is easier than
verifying (B3)+(B4) \cite{Xu-2020-PRA}.

We can check that the definitions of block incoherent state and block
incoherent channel, and also the conditions (B1) to (B4) all return to the
corresponding ones of standard coherence \cite{BCP-2014-PRL,Tong-2016-PRA}.

Now we turn to review the BKB framework of POVM coherence. For a given POVM $%
E=\{E_{j}\}_{j=1}^{n}$, a state $\rho $ is specified as a POVM incoherent
state (with respect to POVM $E$) if \cite{Brub-2019-PRL,Brub-2021-PRA}
\begin{eqnarray}
E_{j}\rho E_{k}=0,j\neq k.  \label{eq2.6}
\end{eqnarray}%
The definition of POVM incoherent channel is related to the canonical
Naimark extension. For the POVM $E=\{E_{j}\}_{j=1}^{n}$ on the $d$%
-dimensional Hilbert space $H,$ introduce an $n$-dimensional Hilbert space $%
H_{\text{R}}$ with $\{|j\rangle \}_{j=1}^{n}$ an orthonormal basis of $H_{%
\text{R}}.$ Write $E=\{E_{j}=A_{j}^{\dagger }A_{j}\}_{j=1}^{n}.$ Notice that
for any unitaries $\{U_{j}\}_{j=1}^{n}$ acting on $\{A_{j}\}_{j=1}^{n}$
respectively, it holds that $E_{j}=(U_{j}A_{j})^{\dagger }(U_{j}A_{j}).$
Since POVM coherence is defined with respect to $E=\{E_{j}\}_{j=1}^{n}$,
then when we express POVM coherence in terms of $\{A_{j}\}_{j=1}^{n}$, the
POVM coherence should be invariant under the unitary transformations $%
\{A_{j}\}_{j=1}^{n}\rightarrow \{U_{j}A_{j}\}_{j=1}^{n}.$ We define the
unitary operator $V$ on $H_{\varepsilon }=H\otimes H_{\text{R}}$ as
\begin{eqnarray}
V &=&\sum_{jk=1}^{n}V_{jk}\otimes |j\rangle \langle k|,  \label{eq2.7}  \\
V_{j1} &=&A_{j},\forall j.   \label{eq2.8}
\end{eqnarray}
For the projective measurement $\overline{P}=\{\overline{P}_{j}\}_{j=1}^{n}$
as
\begin{eqnarray}
\overline{P}_{j}=I_{d}\otimes |j\rangle \langle j|,  \label{eq2.9}
\end{eqnarray}
define the projective measurement $\widetilde{P}=\{\widetilde{P}%
_{j}\}_{j=1}^{n}$ as
\begin{eqnarray}
\widetilde{P}_{j}=V^{\dagger }\overline{P}_{j}V.  \label{eq2.10}
\end{eqnarray}
The projective measurement $\widetilde{P}=\{\widetilde{P}_{j}\}_{j=1}^{n}$
is a canonical Naimark extension of the POVM $E=\{E_{j}\}_{j=1}^{n}$ which
satisfies tr$(E_{j}\rho )=$tr$(\widetilde{P}_{j}\rho \otimes |1\rangle
\langle 1|)$ \cite{Brub-2019-PRL,Brub-2021-PRA}. A channel $\phi $ is
defined as POVM incoherent if $\phi $ allows a Kraus operator decomposition $%
\phi =\{K_{l}\}_{l}$ and there exists a block incoherent channel $\phi
^{\prime }$ with its Block incoherent decomposition $\phi ^{\prime
}=\{K_{l}^{\prime }\}_{l}$ with respect to the projective measurement $%
\widetilde{P}=\{\widetilde{P}_{j}\}_{j=1}^{n}$ on $H_{\varepsilon }=H\otimes
H_{\text{R}}$ such that
\begin{eqnarray}
K_{l}\rho K_{l}^{\dagger }\otimes |1\rangle \langle 1|=K_{l}^{\prime }(\rho
\otimes |1\rangle \langle 1|)K_{l}^{^{\prime }\dagger },\forall l,\forall
\rho .   \label{eq2.11}
\end{eqnarray}
We call such decomposition $\phi =\{K_{l}\}_{l}$ a POVM incoherent
decomposition.

With the definitions of POVM incoherent state and POVM incoherent channel,
the authors in Ref. \cite{Brub-2021-PRA} established the following
conditions that any POVM coherence measure $C(\rho ,E)$ should satisfy.

(P1). Nonnegativity: $C(\rho ,E)\geq 0,$ and $C(\rho ,E)=0$ if and only if $%
\rho $ is POVM incoherent.

(P2). Monotonicity: $C(\phi (\rho ),E)\leq C(\rho ,E)$ for any state $\rho $
and any POVM incoherent channel $\phi .$

(P3). Strong monotonicity: $\sum_{l}$tr$(K_{l}\rho K_{l}^{\dagger })C(\frac{%
K_{l}\rho K_{l}^{\dagger }}{\text{tr}(K_{l}\rho K_{l}^{\dagger })},E)\leq
C(\rho ,E)$ for any state $\rho $ and any POVM incoherent channel $\phi $
with $\phi =\{K_{l}\}_{l}$ a POVM incoherent decomposition.

(P4). Convexity: $C(\sum_{l}p_{l}\rho _{l},E)\leq \sum_{l}p_{l}C(\rho
_{l},E) $ for any states $\{\rho _{l}\}_{l}$ and any probability
distribution $\{p_{l}\}_{l}.$

One can check that when the POVM $E$ is a projective measurement, the definitions of POVM incoherent state and incoherent
channel return to the definitions of block incoherent state and block
incoherent channel, and also (P1)-(P4) return to (B1)-(B4). Since Block
coherence includes standard coherence as special case, then POVM coherence
includes both block coherence and standard coherence as special cases.

There is an efficient method to construct POVM coherence measures as follows.

\emph{Lemma 1} \cite{Brub-2021-PRA}. If $C(\rho _{\varepsilon },\overline{P}%
) $ is a unitarily invariant block coherence measure with respect to $%
\overline{P}$ on $H_{\varepsilon }=H\otimes H_{\text{R}},$ then
\begin{eqnarray}
C(\rho ,E):=C(\sum_{jk=1}^{n}A_{j}\rho A_{k}^{\dagger }\otimes |j\rangle \langle
k|,\overline{P})   \label{eq2.12}
\end{eqnarray}
is a POVM coherence measure, where $\rho _{\varepsilon }$ is any state on $%
H_{\varepsilon },$ unitarily invariant means that $C(U\rho _{\varepsilon
}U^{\dagger },U\overline{P}U^{\dagger })=C(\rho _{\varepsilon },\overline{P}%
) $ for any unitary $U$ on $H_{\varepsilon }.$

\section{POVM coherence via quantum Fisher information}

In this section, we propose a POVM coherence measure based on the quantum
Fisher information (QFI).

QFI is a cornerstone of quantum metrology (a recent review see \cite{Liu-2019-JPA}). For a quantum state $\rho $ and a Hermitian operator $A$ on
$d$-dimensional Hilbert space $H,$ $\rho $ evolves to $\rho (\theta )$
under the unitary dynamics $\rho (\theta )=U\rho U^{\dagger }$ with $U=\exp
(-iA\theta ),$  the QFI of $\rho $ with respect to $A$
can be evaluated as \cite{Braunstein-1994-PRL}
\begin{eqnarray}
F(\rho ,A)=2\sum_{\lambda _{l}+\lambda _{l^{\prime }}>0}\frac{(\lambda
_{l}-\lambda _{l^{\prime }})^{2}}{\lambda _{l}+\lambda _{l^{\prime }}}%
|\langle \varphi_{l}|A|\varphi_{l^{\prime }}\rangle |^{2},   \label{eq3.1}
\end{eqnarray}%
with $\rho =\sum_{l=1}^{d}\lambda _{l}|\varphi_{l}\rangle \langle \varphi_{l}|$ being the
eigendecomposition of $\rho .$ There are many elegant properties for QFI
\cite{Liu-2019-JPA}, we only list three of them we will use in this work.

(F1). $F(U\rho U^{\dagger },UAU^{\dagger })=F(\rho ,A)$ for any unitary $U.$

(F2). $F(\phi (\rho ),A)=F(\rho ,A)$ for any quantum channel $\phi .$

(F3). $F(p\rho _{1}+(1-p)\rho _{2},A)\leq pF(\rho _{1},A)+(1-p)F(\rho
_{2},A) $ for any states $\rho _{1}$, $\rho _{2},$ and $p\in \lbrack 0,1].$

We propose a POVM coherence measure based on QFI as follows.

\emph{Theorem 1.} POVM coherence with respect to the POVM $%
E=\{E_{j}\}_{j=1}^{n}$ based on the quantum Fisher information defined as
\begin{eqnarray}
C_{\text{F}}(\rho ,E)=:\sum_{j=1}^{n}F(\rho ,E_{j})  \label{eq3.2}
\end{eqnarray}%
is a valid POVM coherence measure, where $F(\rho ,E_{j})$ is the quantum
Fisher information of $\rho $ with respect to $E_{j}.$

\emph{Proof.} From (F1) we see that
\begin{eqnarray}
C_{\text{F}}(\rho _{\varepsilon },\overline{P}):=\sum_{j=1}^{n}F(\rho
_{\varepsilon },\overline{P}_{j})  \label{eq3.3}
\end{eqnarray}
is unitarily invariant. Together with Eqs. (\ref{eq2.7},\ref{eq2.8},\ref{eq2.9},\ref{eq2.10}) we have
\begin{eqnarray}
&&F(\sum_{k,k^{\prime }=1}^{n}A_{k}\rho A_{k^{\prime }}^{\dagger }\otimes
|k\rangle \langle k^{\prime }|,\overline{P}_{j})  \notag \\
&=&F(V(\rho \otimes |1\rangle \langle 1|)V^{\dagger },I_{d}\otimes |j\rangle
\langle j|)  \notag \\
&=&F(\rho \otimes |1\rangle \langle 1|,V^{\dagger }(I_{d}\otimes |j\rangle
\langle j|)V)  \notag \\
&=&F(\rho \otimes |1\rangle \langle 1|,\sum_{k,k^{\prime
}=1}^{n}V_{jk}^{\dagger }V_{jk^{\prime }}\otimes |k\rangle \langle k^{\prime
}|)  \notag \\
&=&2\sum_{\lambda_{l}+\lambda_{l^{\prime }}>0}\frac{(\lambda_{l}-\lambda_{l^{\prime }})^{2}}{%
\lambda_{l}+\lambda_{l^{\prime }}}  \notag \\
&&\ \ \ \ \ \cdot |\langle \varphi_{l}|\langle 1|(\sum_{k,k^{\prime
}=1}^{n}V_{jk}^{\dagger }V_{jk^{\prime }}\otimes |k\rangle \langle k^{\prime
}|)|\varphi_{l^{\prime }}\rangle |1\rangle |^{2}  \notag \\
&=&2\sum_{\lambda_{l}+\lambda_{l^{\prime }}>0}\frac{(\lambda_{l}-\lambda_{l^{\prime }})^{2}}{%
\lambda_{l}+\lambda_{l^{\prime }}}|\langle \varphi_{l}|A_{j}^{\dagger }A_{j}|\varphi_{l^{\prime }}\rangle
|^{2}  \notag \\
&=&F(\rho ,E_{j}),  \notag
\end{eqnarray}%
where $\rho =\sum_{l=1}^{d}\lambda_{l}|\varphi_{l}\rangle \langle \varphi_{l}|$ is the eigendecomposition of
$\rho .$ Then we get $C_{\text{F}}(\rho _{\varepsilon },\overline{P})=C_{%
\text{F}}(\rho ,E)$.

Applying lemma 1, we now prove that $C_{\text{F}}(\rho _{\varepsilon },%
\overline{P})$ is a valid block coherence measure. We see, (F2) implies that $C_{%
\text{F}}(\rho _{\varepsilon },\overline{P})$ satisfies (B2), (F3) implies
that $C_{\text{F}}(\rho _{\varepsilon },\overline{P})$ satisfies (B4). To
prove $C_{\text{F}}(\rho _{\varepsilon },\overline{P})$ satisfies (B5),
suppose $\rho _{\varepsilon }=p\rho _{\varepsilon 1}\oplus (1-p)\rho
_{\varepsilon 2}$ as described in (B5). let $\rho _{\varepsilon
1}=\sum_{l}\lambda _{1,l}|\varphi _{1,l}\rangle \langle \varphi _{1,l}|$ be
the eigendecomposition of $\rho _{\varepsilon 1},$ $\rho _{\varepsilon
2}=\sum_{l}\lambda _{2,l}|\varphi _{2,l}\rangle \langle \varphi _{2,l}|$ be
the eigendecomposition of $\rho _{\varepsilon 2}.$ From Eq. (\ref{eq3.1}) we have
\begin{eqnarray}
&&F(p_{1}\rho _{\varepsilon 1}\oplus p_{2}\rho _{\varepsilon 2},%
\overline{P}_{j})  \notag \\
&=&p_{1}\sum_{\lambda _{1,l}+\lambda _{1,l^{\prime }}>0}\frac{(\lambda
_{1,l}-\lambda _{1,l^{\prime }})^{2}}{\lambda _{1,l}+\lambda _{1,l^{\prime }}%
}|\langle \varphi _{1,l}|\overline{P}_{j}|\varphi _{1,l^{\prime }}\rangle
|^{2}  \notag \\
&&+p_{2}\sum_{\lambda _{2,l}+\lambda _{2,l^{\prime }}>0}\frac{(\lambda
_{2,l}-\lambda _{2,l^{\prime }})^{2}}{\lambda _{2,l}+\lambda _{2,l^{\prime }}%
}|\langle \varphi _{2,l}|\overline{P}_{j}|\varphi _{2,l^{\prime }}\rangle
|^{2}    \notag \\
&=&p_{1}F(\rho _{\varepsilon 1},\overline{P}_{j})+p_{2}F(\rho
_{\varepsilon 2},\overline{P}_{j}).  \notag
\end{eqnarray}
Summing over $j,$ we get that $C_{\text{F}}(\rho _{\varepsilon },\overline{P}%
)$ satisfies (B5). \

We then only need to prove that $C_{\text{F}}(\rho _{\varepsilon },\overline{P})$
satisfies (B1). To this end, we prove Lemma 2 below.

\emph{Lemma 2.} For the projective measurement $P=\{P_{j}\}_{j=1}^{n}$ on
the $d$-dimensional Hilbert space $H,$
\begin{eqnarray}
C_{\text{F}}(\rho ,P) &:&=\sum_{j=1}^{n}F(\rho ,P_{j})=0  \label{eq3.4} \\
&\Leftrightarrow &P_{j}\rho P_{k}=0,\forall j\neq k.   \label{eq3.5}
\end{eqnarray}

\emph{Proof of Lemma 2.} Suppose $C_{\text{F}}(\rho ,P)=0.$ Since $F(\rho ,P_{j})\geq 0$ for any $j,$ then $C_{%
\text{F}}(\rho ,P)=0$ if and only if $F(\rho ,P_{j})=0$ for any $j.$ Notice
that $P_{j}\rho P_{k}=0$ for any $j\neq k$ is equivalent to say that $\rho $
is of the form $\rho =\oplus _{j=1}^{n}p_{j}\rho _{j}$ with $\{p_{j}\}_{j}$ a
probability distribution and $\rho _{j}=P_{j}\rho _{j}P_{j}$ for any $j.$

\ Let $P_{j}=\sum_{\alpha=1}^{d_{j}}|\phi _{j\alpha}\rangle \langle \phi
_{j\alpha}|$ be an eigendecomposition of $P_{j}$, then $\{|\phi
_{j\alpha}\rangle \}_{j\alpha}$ constitute an orthonormal basis of $H$ and $%
\sum_{j=1}^{n}d_{j}=d.$  We relabel the orthonormal basis $\{|\phi _{j\alpha}\rangle
\}_{j\alpha} $ as $\{|\phi _{j\alpha}\rangle
\}_{j\alpha}=\{|\phi _{11}\rangle ,|\phi _{12}\rangle ,...,|\phi
_{1d_{1}}\rangle ,|\phi _{21}\rangle ,|\phi _{22}\rangle ,...,|\phi
_{2d_{2}}\rangle ,...\}=\{|1\rangle ,|2\rangle ,...,|d_{1}\rangle ,|d_{1}+1\rangle
,|d_{1}+2\rangle ,...,|d_{1}+d_{2}\rangle ,...\}=\{|m\rangle \}_{m=1}^{d}.$

let
\begin{eqnarray}
\rho =\sum_{l=1}^{d}\lambda _{l}|\varphi _{l}\rangle \langle \varphi _{l}|  \label{eq3.6}
\end{eqnarray}
be the eigendecomposition of $\rho ,$ with $\lambda _{l}\geq 0$ for any $l$
and $\sum_{l=1}^{d}\lambda _{l}=1.$ Without loss of generality, we always
assume that $\lambda _{1}\geq \lambda _{2}\geq ...\geq \lambda _{d}.$ We
express $\{|\varphi _{l}\rangle \}_{l=1}^{d}$ in $\{|m\rangle \}_{m=1}^{d}$
as
\begin{equation}
|\varphi _{l}\rangle =\sum_{m=1}^{d}\varphi _{lm}|m\rangle .  \label{eq3.7}
\end{equation}%
Introduce a $d\times d$ matrix
\begin{eqnarray}
\Phi &=&\left( \varphi _{lm}\right) _{1\leq l\leq d;1\leq m\leq d} \nonumber \\
&=&\left(
\begin{array}{cccc}
\varphi _{11} & \varphi _{12} & ... & \varphi _{1d} \\
\varphi _{21} & \varphi _{22} & ... & \varphi _{2d} \\
... & ... & ... & ... \\
\varphi _{d1} & \varphi _{d2} & ... & \varphi _{dd}%
\end{array}%
\right)   \label{eq3.8}
\end{eqnarray}%
We set three steps to consider different situations of $\{\lambda
_{l}\}_{l=1}^{d}.$

(1). Situation 1: $\lambda _{1}>\lambda _{2}>...>\lambda _{d}.$ In this
situation, $\{|\varphi _{l}\rangle \}_{l=1}^{d}$ forms an orthonormal basis
of $H,$ $\Phi $ is a unitary matrix, from Eq. (\ref{eq3.1}), $F(\rho ,P_{1})=0$ reads
\begin{eqnarray}
\sum_{\alpha =1}^{d_{1}}\langle \varphi _{l}|\phi _{1\alpha }\rangle \langle
\phi _{1\alpha }|\varphi _{l^{\prime }}\rangle =0,\forall l\neq l^{\prime },  \label{eq3.9}
\end{eqnarray}
that is
\begin{eqnarray}
\sum_{m=1}^{d_{1}}\varphi _{lm}^{\ast }\varphi _{l^{\prime }m}=0,\forall
l\neq l^{\prime }.  \label{eq3.10}
\end{eqnarray}

Consider the $d\times d_{1}$ matrix $\left( \varphi _{lm}\right) _{1\leq
l\leq d;1\leq m\leq d_{1}},$ since $d_{1}$ column vectors $\{\left( \varphi
_{lm}\right) _{1\leq l\leq d}\}_{m=1}^{d_{1}}$ are orthogonal to each other,
then the column rank
\begin{eqnarray}
\text{rank}(\{\left( \varphi _{lm}\right) _{1\leq l\leq
d}\}_{m=1}^{d_{1}})=d_{1}.  \label{eq3.11}
\end{eqnarray}%
We know that for any matrix, the rank of column vectors equals the rank of
row vectors, thus the row rank
\begin{eqnarray}
\text{rank}(\{\left( \varphi _{lm}\right) _{1\leq m\leq
d_{1}}\}_{l=1}^{d})=d_{1}.  \label{eq3.12}
\end{eqnarray}%
Eq. (\ref{eq3.10}) says $d$ row vectors $\{\left( \varphi _{lm}\right) _{1\leq m\leq
d_{1}}\}_{l=1}^{d}$ are orthogonal to each other, then there must be $d_{1}$
row vectors being nonzero (not all elements are zero), and other $d-d_{1}$
row vectors being zero (all elements are zero). let $\{\left( \varphi
_{l_{1},m}\right) _{1\leq m\leq d_{1}}\}_{l_{1}}$ denotes the $d_{1}$
nonzero rows in $\{\left( \varphi _{lm}\right) _{1\leq m\leq
d_{1}}\}_{l=1}^{d},$ that is, the set $\{l_{1}\}_{l_{1}}$ has just $d_{1}$
distinct numbers in $\{l\}_{l=1}^{d},$ we write the number of $%
\{l_{1}\}_{l_{1}}$ as $|\{l_{1}\}_{l_{1}}|=d_{1}.$

Similarly, $F(\rho ,P_{j})$ yields
\begin{eqnarray}
|\{l_{j}\}_{l_{j}}|=d_{j}.  \label{eq3.13}
\end{eqnarray}
Since $\Phi =\left( \varphi _{lm}\right) _{1\leq l\leq d;1\leq m\leq d}$ is
a unitary matrix, then rank$\Phi =d,$ here rank$\Phi $ is the rank of column
vectors $\{\left( \varphi _{lm}\right) _{1\leq l\leq d}\}_{m=1}^{d}$ and
also the row vectors $\{\left( \varphi _{lm}\right) _{1\leq m\leq
d}\}_{l=1}^{d}.$ This fact implies that
\begin{eqnarray}
\{l_{j}\}_{l_{j}}\cap \{l_{k}\}_{l_{k}}=\varnothing ,\forall j\neq k,  \label{eq3.14}
\end{eqnarray}%
otherwise there must be a row vector in $\{\left( \varphi _{lm}\right)
_{1\leq m\leq d}\}_{l=1}^{d}$ being zero vector, this contradicts rank$\Phi
=d.$
Consequently, each $|\varphi _{l}\rangle $ of $\{|\varphi _{l}\rangle
\}_{l=1}^{d}$ is just in the range of one $P_{j},$
\begin{eqnarray}
|\varphi _{l}\rangle \subset \text{Range}\{P_{j}\},  \label{eq3.15}
\end{eqnarray}
then $\rho $ must be of the form $\rho =\oplus _{j=1}^{n}p_{j}\rho _{j}$ with $\{p_{j}\}_{j}$ a
probability distribution and $\rho _{j}=P_{j}\rho _{j}P_{j}$ for any $j.$

(2). Situation 2: $\lambda _{1}=\lambda _{2}=...=\lambda _{D_{1}}>\lambda
_{D_{1}+1}=\lambda _{D_{1}+2}=...=\lambda _{D_{1}+D_{2}}>\lambda
_{D_{1}+D_{2}+1}=...=\lambda _{D_{1}+D_{2}+...+D_{N}}\geq 0,$ with $%
d=D_{1}+D_{2}+...+D_{N}.$ We also let $D_{0}=0.$ For this situation, we rewrite Eq. (\ref{eq3.6}) as
\begin{eqnarray}
\rho=\lambda _{1}\sum_{l=1}^{D_{1}}|\varphi _{l}\rangle \langle \varphi
_{l}|+\lambda _{D_{1}+1}\sum_{l=D_{1}+1}^{D_{1}+D_{2}}|\varphi _{l}\rangle
\langle \varphi _{l}|+...;  \label{eq3.16}
\end{eqnarray}
and rewrite Eq. (\ref{eq3.8}) as a block matrix
\begin{eqnarray}
\Phi &=&\left( \varphi _{lm}\right) _{1\leq l\leq d;1\leq m\leq d} \nonumber \\
&=&\left(
\begin{array}{cccc}
\Phi _{11} & \Phi _{12} & ... & \Phi _{1n} \\
\Phi _{21} & \Phi _{22} & ... & \Phi _{2n} \\
... & ... & ... & ... \\
\Phi _{N1} & \Phi _{N2} & ... & \Phi _{Nn}%
\end{array}%
\right)   \label{eq3.17}
\end{eqnarray}
with the block elements for example
\begin{eqnarray}
\Phi _{11} &=&\left( \varphi _{lm}\right) _{1\leq l\leq D_{1};1\leq m\leq
d_{1}} \nonumber \\
&=&\left(
\begin{array}{cccc}
\varphi _{11} & \varphi _{12} & ... & \varphi _{1d_{1}} \\
\varphi _{21} & \varphi _{22} & ... & \varphi _{2d_{1}} \\
... & ... & ... & ... \\
\varphi _{D_{1}1} & \varphi _{D_{2}2} & ... & \varphi _{D_{1}d_{1}}%
\end{array}%
\right).    \label{eq3.18}
\end{eqnarray}%
From Eq. (\ref{eq3.1}), we see that $F(\rho ,P_{1})=0$ if and
only if
\begin{equation}
\Phi _{l1}\Phi _{l^{\prime }1}^{\dagger }=0,\forall l\neq l^{\prime }.  \label{eq3.19}
\end{equation}
Notice that if $P_{j}=\sum_{\alpha =1}^{d_{j}}|\phi _{j\alpha }\rangle
\langle \phi _{j\alpha }|$ is an eigendecomposition of $P_{j},$ then so is
\begin{eqnarray}
P_{j}=\sum_{\alpha =1}^{d_{j}}U_{j}^{*}|\phi _{j\alpha }\rangle \langle \phi
_{j\alpha }|U_{j}^{t}   \label{eq3.20}
\end{eqnarray}
 for any $d_{j}\times d_{j}$ unitary $U_{j},$  where $U_{j}^{*}$ stands for the conjugate of $U_{j}$, $U_{j}^{t}$ stands for the transpose of $U_{j}$, and $U_{j}^{\dag}=U_{j}^{*t}$.
Similarly, Eq. (\ref{eq3.6}) can be rewritten as
\begin{eqnarray}
\rho &=&\lambda _{1}\sum_{l=1}^{D_{1}}W_{1}^{t}|\varphi _{l}\rangle \langle
\varphi _{l}|W_{1}^{*} \nonumber \\
&&+\lambda
_{D_{1}+1}\sum_{l=D_{1}+1}^{D_{1}+D_{2}}W_{2}^{t}|\varphi _{l}\rangle \langle
\varphi _{l}|W_{2}^{*}+...  \label{eq3.21}
\end{eqnarray}
for any $\{W_{k}\}_{k=1}^{N}$ with each $W_{k}$ a $D_{k}\times D_{k}$
unitary matrix. Under the definition of Eq. (\ref{eq3.8}), Eqs. (\ref{eq3.20}, \ref{eq3.21}) correspond to
\begin{eqnarray}
&&\Phi ^{\prime }=(\oplus _{k=1}^{N}W_{k})\Phi (\oplus _{j=1}^{n}U_{j}) \nonumber \\
&=&\left(
\begin{array}{cccc}
W_{1}\Phi _{11}U_{1} & W_{1}\Phi _{12}U_{2} & ... & W_{1}\Phi _{1n}U_{n} \\
W_{2}\Phi _{21}U_{1} & W_{2}\Phi _{22}U_{2} & ... & W_{2}\Phi _{2n}U_{n} \\
... & ... & ... & ... \\
W_{N}\Phi _{N1}U_{1} & W_{N}\Phi _{N2}U_{2} & ... & W_{N}\Phi _{Nn}U_{n}%
\end{array}%
\right), \ \  \label{eq3.22}
\end{eqnarray}
which is a unitary matrix, we denote it by $\Phi ^{\prime }=\left( \varphi
_{lm}^{\prime }\right) _{1\leq l\leq d;1\leq m\leq d}=\left( \Phi
_{kj}^{\prime }\right) _{1\leq k\leq N;1\leq j\leq n}.$ Below we will use
the notation $\Phi ^{(l)}=\left( \varphi _{lm}^{(l)}\right) _{1\leq l\leq
d;1\leq m\leq d}$ $=\left( \Phi _{kj}^{(l)}\right) _{1\leq k\leq N;1\leq
j\leq n}$ similarly defined. Notice that rank$\Phi _{kj}=$rank$(W_{k}\Phi
_{kj}U_{j}).$

Suppose $\{$rank$\Phi _{k1}=r_{k}\}_{k=1}^{N},$ since the row rank equals
the column rank, then $\sum_{k=1}^{N}r_{k}=d_{1}.$ If rank$\Phi
_{11}=r_{1}>1,$ using the sigular value decomposition, there exist unitary $%
\{W_{1}^{(1)}\bigskip ,U_{1}^{(1)}\}$ such that
\begin{eqnarray}
W_{1}^{(1)}\Phi _{11}U_{1}^{(1)}
=\left(
\begin{array}{cccccc}
c_{1} & 0 & ... & 0 & 0 & ... \\
0 & c_{2} & ... & 0 & 0 & ... \\
... & ... & ... & ... & 0 & ... \\
0 & 0 & ... & c_{r_{1}} & 0 & ... \\
0 & 0 & 0 & 0 & 0 & ... \\
... & ... & ... & ... & ... & ...%
\end{array}%
\right) .   \label{eq3.23}
\end{eqnarray}%
For such $\{W_{1}^{(1)},U_{1}^{(1)}\},$ and other unitary matrices $%
\{W_{k}^{(1)}\}_{k=2}^{N},\{U_{j}^{(1)}\}_{j=2}^{n}$ being identity
matrices with corresponding dimensions, now consider the matrix $\Phi
^{(1)}=(\oplus _{k=1}^{N}W_{k}^{(1)})\Phi (\oplus _{j=1}^{n}U_{j}^{(1)}).$
From $\Phi _{l1}^{(1)}\Phi _{l^{\prime }1}^{(1)\dagger }=0,\forall l\neq
l^{\prime },$ we get $\{\varphi _{lm}^{(1)}=0\}_{D_{1}+1\leq l\leq d;1\leq
m\leq r_{1}}.$ Since $\Phi ^{(1)}$ is unitary, each row or each column has
the unit length, then $|c_{1}|=|c_{2}|=...=|c_{r_{1}}|=1,$ $\{\varphi
_{lm}^{(1)}=0\}_{1\leq l\leq r_{1};d_{1}+1\leq m\leq d}.$ Thus
\begin{eqnarray}
\{(\varphi _{lm}^{(1)})_{1\leq m\leq d}\}_{1\leq l\leq r_{1}}\subset \text{%
Range}\{P_{1}\}.  \label{eq3.24}
\end{eqnarray}%
If rank$\Phi _{11}=r_{1}=0,$ we skip this step, or equivalently, we define $%
\Phi ^{(1)}=(\oplus _{k=1}^{N}W_{k}^{(1)})\Phi (\oplus
_{j=1}^{n}U_{j}^{(1)})=\Phi $ with  $\{W_{k}^{(1)}\}_{k=1}^{N},\{U_{j}^{(1)}%
\}_{j=1}^{n}$ all identity matrices.

If rank$\Phi _{21}^{(1)}=r_{2}>0,$ we define $\Phi ^{(2)}=(\oplus
_{k=1}^{N}W_{k}^{(2)})\Phi ^{(1)}(\oplus _{j=1}^{n}U_{j}^{(2)})$ as follows. Using
the sigular value decomposition, there exist unitary $%
\{W_{2}^{(2)},U_{1}^{(2)}=I_{r_{1}}\oplus u_{1}^{(2)}\}$ with $I_{r_{1}}$
the $r_{1}\times r_{1}$ identity matrix and $u_{1}^{(2)}$ a $%
(d_{1}-r_{1})\times (d_{1}-r_{1})$ unitary matrix, such that
\begin{eqnarray}
W_{2}^{(2)}\Phi _{21}^{(1)}U_{1}^{(2)}= \ \ \ \ \ \ \ \ \ \ \  \ \ \ \ \ \ \ \ \ \ \  \ \ \ \ \ \ \ \ \ \ \  \ \ \ \ \ \ \ \ \ \ \ \ \ \   \nonumber \\
\left(
\begin{array}{cccccccc}
0_{D_{1}+1,1} & ... & 0_{D_{1}+1,r_{1}} & c_{r_{1}+1} & ... & 0 & 0 & ... \\
... & ... & ... & 0 & ... & 0 & 0 & ... \\
0_{D_{1}+r_{2},1} & ... & 0_{D_{1}+r_{2},r_{1}} & 0 & ... & c_{r_{1}+r_{2}}
& 0 & ... \\
0 & 0 & 0 & 0 & ... & 0 & 0 & ... \\
... & ... & ... & ... & ... & ... & ... & ...%
\end{array}%
\right) ,   \nonumber \\ \label{eq3.25}
\end{eqnarray}
where $|c_{r_{1}+1}|=|c_{r_{1}+2}|=...=|c_{r_{1}+r_{2}}|=1,$ $%
0_{D_{1}+r_{2},r_{1}}$ means $\varphi _{D_{1}+r_{2},r_{1}}^{(2)}=0,$ $%
\{W_{1}^{(2)}\}\cup \{W_{k}^{(2)}\}_{k=3}^{N},\{U_{j}^{(2)}\}_{j=2}^{n}$ are
identity matrices with corresponding dimensions. If rank$\Phi
_{21}^{(1)}=r_{2}=0,$ we skip this step, or equivalently, we define $\Phi
^{(2)}=(\oplus _{k=1}^{N}W_{k}^{(2)})\Phi ^{(1)}(\oplus
_{j=1}^{n}U_{j}^{(2)})=\Phi ^{(1)}$ with $\{W_{k}^{(2)}\}_{k=1}^{N},%
\{U_{j}^{(2)}\}_{j=1}^{n}$ all identity matrices.

Repeat this procedure we will get
\begin{eqnarray}
\Phi ^{(N)} &=&(\oplus _{k=1}^{N}W_{k}^{(N-1)})\Phi ^{(N-1)}(\oplus
_{j=1}^{n}U_{j}^{(N-1)}) \nonumber \\
&=&(\oplus _{k=1}^{N}W_{k})\Phi (\oplus _{j=1}^{n}U_{j}),  \label{eq3.26} \\
W_{k} &=&W_{k}^{(N-1)}...W_{k}^{(2)}W_{k}^{(1)},  \label{eq3.27} \\
U_{j} &=&U_{j}^{(1)}U_{j}^{(2)}...U_{j}^{(N-1)}.  \label{eq3.28+}
\end{eqnarray}
In the first $d_{1}$ columns of unitary matrix $\Phi ^{(N)},$ there are only $%
d_{1}$ nonzero elements $\{c_{j}\}_{j=1}^{d_{1}}$ each of them having
modulus $1,$ each column has just one of them. In unitary matrix $\Phi
^{(N)},$ for any $c_{j}\in \{c_{j}\}_{j=1}^{d_{1}},$ the row of containing $c_{j}$
has just one nonzero element $c_{j}.$ We denote the set of $d_{1}$
rows in $\Phi ^{(N)}$ each just has one of $\{c_{j}\}_{j=1}^{d_{1}}$
as $\Psi _{1}=\{(\varphi _{lm}^{(N)})_{1\leq m\leq d}:D_{k}+1\leq l\leq
D_{k}+r_{k},k=0,1,...,N-1\},$ clearly,
\begin{eqnarray}
\Psi _{1} &\subset &\text{Range}\{P_{1}\},  \label{eq3.29+} \\
|\Psi _{1}| &=&d_{1},  \label{eq3.30+}
\end{eqnarray}
where $|\Psi _{1}|$ denote the number of vectors in $\Psi _{1}.$

In unitary matrix $\Phi ^{(N)},$ delete the rows and columns containing the elements $%
\{c_{j}\}_{j=1}^{d_{1}}$ above, the remained matrix is a $(d-d_{1})\times
(d-d_{1})$ unitary matrix with all row vectors belonging to Range$%
\{I_{d}-P_{1}\}.$ This remained matrix corresponds to the remained state
\begin{eqnarray}
\rho _{1}&=&\rho -\lambda _{1}\sum_{l=1}^{r_{1}}W_{1}^{t}|\varphi _{l}\rangle
\langle \varphi _{l}|W_{1}^{*} \nonumber \\
&&-\lambda
_{D_{1}+1}\sum_{l=D_{1}+1}^{D_{1}+r_{2}}W_{2}^{t}|\varphi _{l}\rangle \langle
\varphi _{l}|W_{2}^{*}-...,  \label{eq3.31+}
\end{eqnarray}%
then we taken out the part of $\rho$ projecting into the subspace of Range$\{P_{1}\}$. 
Now repeat the procedure for $P_{2}$, $P_{3},...$, $P_{n}$,  we will get that there exists an
eigendecomposition $\rho =\sum_{l=1}^{d}\lambda _{l}|\varphi _{l}\rangle
\langle \varphi _{l}|$ such that each $|\varphi _{l}\rangle $ just belongs
to one Range$\{P_{j}\}.$ 

Combine situations 1-2 above, we then end the proof of lemma 2. Having lemma 2, we immediately get that $C_{\text{F}}(\rho _{\varepsilon },%
\overline{P})$ satisfies (B1), and then end the proof of Theorem 1.

As a special case of Theorem 1, when POVM $E$ is a rank-1 projective measurement, Theorem 1 becomes the Corollary 1 below.

\emph{Corollary 1.} Suppose $\{|j\rangle \}_{j=1}^{d}$ is an orthonormal
basis of $d$-dimensional Hilbert space, then
\begin{eqnarray}
C_{\text{F}}(\rho ,\{|j\rangle \}_{j=1}^{d})=\sum_{j=1}^{d}F(\rho ,|j\rangle
\langle j|)  \label{eq3.28}
\end{eqnarray}
is a valid standard coherence measure under BCP framework with respect to $\{|j\rangle \}_{j=1}^{d}.$

Remark that in Ref. \cite{Li-PRA-2021}, the authors claimed that $C_{\text{F}}(\rho ,\{|j\rangle \}_{j=1}^{d})$ is a valid standard coherence
measure (which coincides with Corollary 1 above, up to a factor 1/4), and proved this assertion by proving $C_{\text{F}}(\rho
,\{|j\rangle \}_{j=1}^{d})=C_{\text{F}}^{C}(\rho ,\{|j\rangle \}_{j=1}^{d})$
in Proposition 4 with $C_{\text{F}}^{C}(\rho ,\{|j\rangle \}_{j=1}^{d})$
defined in definition 3. In appendix A we will show that that proof is not
justified, i.e., $C_{\text{F}}(\rho ,\{|j\rangle \}_{j=1}^{d})=C_{%
\text{F}}^{C}(\rho ,\{|j\rangle \}_{j=1}^{d})$ is not true in general.

\section{An application of POVM coherence via QFI}
We provide an application of $C_{\text{F}}(\rho ,E)$ in quantum metrology.
QFI gives the ultimate precision bound on the estimation of parameters
encoded in a quantum state. Assuming that we start from an initial state $%
\rho $ and $A$ is a Hermitian operator, $\rho $ evolves to $\rho (\theta )$
under the unitary dynamics $\rho (\theta )=U\rho U^{\dagger }$ with $U=\exp
(-iA\theta )$. QFI $F(\rho ,A)$ constrains the achievable precision in
estimation of the parameter $\theta $ via the quantum Cram\'{e}r-Rao bound
as
\begin{eqnarray}
(\Delta \theta )^{2}\geq \frac{1}{NF(\rho ,A)},  \label{eq4.1}
\end{eqnarray}%
where $(\Delta \theta )^{2}$ is the variance of $\theta $, $N$ is the number
of independent repetitions.

Let $U=\exp (-iE_{j}\theta _{j})$, Eqs. (\ref{eq4.1},\ref{eq3.2}) yield
\begin{eqnarray}
(\Delta \theta _{j})^{2} &\geq &\frac{1}{NF(\rho ,E_{j})},  \label{eq4.2} \\
\frac{1}{(\Delta \theta _{j})^{2}} &\leq &NF(\rho ,E_{j}),  \label{eq4.3} \\
\sum_{j=1}^{n}\frac{1}{(\Delta \theta _{j})^{2}} &\leq &NC_{F}(\rho ,E).  \label{eq4.4}
\end{eqnarray}%
Eq. (\ref{eq4.4}) sets an uncertainty relation on $\{(\Delta \theta
_{j})^{2}\}_{j=1}^{n}$ by the upper bound $NC_{F}(\rho ,E).$

Remark that when $E$ is a rank-1 projective measurement, Eq. (\ref{eq4.4}) recovers the corresponding results in \cite{Li-PRA-2021}.

\section{Summary}

We proposed a POVM coherence measure via QFI under the BKB framework and
provide an application in estimation theory. QFI is a core concept in
 quantum metrology, we hope that this POVM coherence measure will provide new
insights in quantum metrology and in some quantum information processings.

\section*{ACKNOWLEDGMENTS}
This work was supported by the Chinese Universities Scientific Fund under Grant No. 2452021067. The author thanks Lin Zhang and Jing Liu for helpful discussions.

\section*{Appendix: Disproof of $C_{\text{F}}(\protect\rho %
,\{|j\rangle \}_{j=1}^{d})=C_{\text{F}}^{C}(\protect\rho ,\{|j\rangle
\}_{j=1}^{d})$}

\setcounter{equation}{0} \renewcommand%
\theequation{A\arabic{equation}}
In this section, we disprove the result $C_{\text{F}}(\protect\rho %
,\{|j\rangle \}_{j=1}^{d})=C_{\text{F}}^{C}(\protect\rho ,\{|j\rangle
\}_{j=1}^{d})$ claimed in Proposition 4 of Ref. \cite{Li-PRA-2021}.

Suppose $\{|j\rangle \}_{j=1}^{d}$ is an orthonormal basis of a $d$%
-dimensional Hilbert space $H.$ For state $\rho ,$ $C_{\text{F}}(\rho
,\{|j\rangle \}_{j=1}^{d})$ and $C_{\text{F}}^{C}(\rho ,\{|j\rangle
\}_{j=1}^{d})$ are defined as \cite{Li-PRA-2021} (in \cite{Li-PRA-2021} the authors defined $C_{\text{F}}(\rho ,\{|j\rangle \}_{j=1}^{d})=\frac{1}{4}C_{\text{F}}(\rho,\{|j\rangle \}_{j=1}^{d})$, we omit the factor $\frac{1}{4}$)
\begin{eqnarray}
C_{\text{F}}(\rho ,\{|j\rangle \}_{j=1}^{d}) &=&\sum_{j=1}^{d}F(\rho
,|j\rangle \langle j|),   \label{eqA.1} \\
C_{\text{F}}^{C}(\rho ,\{|j\rangle \}_{j=1}^{d}) &=&\min_{\{p_{k},|\psi
_{k}\rangle \}_{k}}\sum_{k}p_{k}C_{\text{F}}(|\psi _{k}\rangle ,\{|j\rangle
\}_{j=1}^{d}),  \label{eqA.2} \nonumber \\
\end{eqnarray}%
where min runs over all pure state ensembles $\{p_{k},|\psi _{k}\rangle
\}_{k}$ of $\rho ,$ i.e., $\rho =$ $\sum_{k}p_{k}|\psi _{k}\rangle \langle
\psi _{k}|$ with $\{p_{k}\}_{k}$ a probability distribution and $\{|\psi
_{k}\rangle \}_{k}$ normalized pure states. Since $F(\rho ,A)$ is convex then
\begin{eqnarray}
F(\rho ,|j\rangle ) &\leq &\sum_{k}p_{k}F(|\psi _{k}\rangle ,|j\rangle),\forall j,  \label{eqA.3} \\
C_{\text{F}}(\rho ,\{|j\rangle \}_{j=1}^{d}) &\leq &C_{\text{F}}^{C}(\rho ,\{|j\rangle \}_{j=1}^{d}),  \label{eqA.4}
\end{eqnarray}
and the question remained is whether the equality can be achieved.

We would like to consider the more general case of $C_{\text{F}}(\rho ,E)$
and $C_{\text{F}}^{C}(\rho ,E)$ with respect to the POVM $%
E=\{E_{j}\}_{j=1}^{n}$ as
\begin{eqnarray}
C_{\text{F}}(\rho ,E) &=&\sum_{j=1}^{n}F(\rho ,E_{j}),  \label{eqA.5}  \\
C_{\text{F}}^{C}(\rho ,E) &=&\min_{\{p_{k},|\psi _{k}\rangle
\}_{k}}\sum_{k}p_{k}C_{\text{F}}(|\psi _{k}\rangle ,E).  \label{eqA.6}
\end{eqnarray}
Since $F(\rho ,A)$ is convex then
\begin{eqnarray}
F(\rho ,E_{j}) &\leq &\sum_{k}p_{k}F(|\psi _{k}\rangle ,E_{j}),\forall j,  \label{eqA.7}  \\
C_{\text{F}}(\rho ,E) &\leq &C_{\text{F}}^{C}(\rho ,E),  \label{eqA.8}
\end{eqnarray}
and $C_{\text{F}}(\rho ,E)=C_{\text{F}}^{C}(\rho ,E)$ if and only if there
exists a pure state ensemble $\{p_{k},|\psi _{k}\rangle \}_{k}$ such that $%
F(\rho ,E_{j})=\sum_{k}p_{k}F(|\psi _{k}\rangle ,E_{j})$ for any $j.$ When $%
\rho $ is a pure state $|\psi \rangle ,$ $C_{\text{F}}(|\psi \rangle ,E)=C_{%
\text{F}}^{C}(|\psi \rangle ,E)$ evidently holds, then in the following we
mainly discuss the case of mixed state $\rho .$ Note that for pure state $%
|\psi _{k}\rangle ,$
\begin{eqnarray}
F(|\psi _{k}\rangle ,E_{j})=\langle \psi _{k}|E_{j}^{2}|\psi _{k}\rangle
-\langle \psi _{k}|E_{j}|\psi _{k}\rangle ^{2},  \label{eqA.9}
\end{eqnarray}
that is, $F(|\psi _{k}\rangle ,E_{j})$ equals the variance of $E_{j}$
with respect to $|\psi _{k}\rangle .$

As in Ref. \cite{Yu-2013-arxiv}, we write $F(\rho ,A)$ in Eq. (\ref{eq3.1}) as
\begin{eqnarray}
F(\rho ,A)=\text{tr}(\rho A^{2})-\text{tr}(Z_{A}^{2}),  \label{eqA.10}
\end{eqnarray}
where
\begin{eqnarray}
Z_{A}=\sum_{l,l^{\prime }=1}^{d}\sqrt{\frac{2\lambda _{l}\lambda _{l^{\prime
}}}{\lambda _{l}+\lambda _{l^{\prime }}}}|\varphi _{l}\rangle \langle
\varphi _{l}|A|\varphi _{l^{\prime }}\rangle \langle \varphi _{l^{\prime }}|,  \label{eqA.11}
\end{eqnarray}
with $\rho=\sum_{l=1}^{d}\lambda _{l}|\varphi _{l}\rangle \langle \varphi
_{l}|$ the eigendecomposition of $\rho .$ In Eq. (\ref{eqA.11}), when $\lambda
_{l}=\lambda _{l^{\prime }}=0,$ we define $\sqrt{\frac{2\lambda _{l}\lambda
_{l^{\prime }}}{\lambda _{l}+\lambda _{l^{\prime }}}}=0.$ For the pure state
ensemble $\rho =$ $\sum_{k}p_{k}|\psi _{k}\rangle \langle \psi _{k}|,$
\begin{eqnarray}
&&\sum_{k}p_{k}F(|\psi _{k}\rangle ,A) \nonumber \\
&=&\sum_{k}p_{k}(\langle \psi
_{k}|A^{2}|\psi _{k}\rangle -\langle \psi _{k}|A|\psi _{k}\rangle ^{2}) \nonumber \\
&=&\text{tr}(\rho A^{2})-\sum_{k}p_{k}\langle \psi _{k}|A|\psi _{k}\rangle
^{2}. \label{eqA.12}
\end{eqnarray}
Since $F(\rho ,A)$ is convex, then
\begin{eqnarray}
F(\rho ,A)\leq \sum_{k}p_{k}F(|\psi _{k}\rangle ,A), \label{eqA.13}
\end{eqnarray}
this is equivalent to
\begin{eqnarray}
\sum_{k}p_{k}\langle \psi _{k}|A|\psi _{k}\rangle ^{2}\leq \text{tr}%
(Z_{A}^{2}). \label{eqA.14}
\end{eqnarray}
We now discuss which pure state ensemble $\{p_{k},|\psi _{k}\rangle \}_{k}$
can achieve the equality above.

Any pure state ensemble $\{p_{k},|\psi _{k}\rangle \}_{k=1}^{d^{\prime }}$
can be generated by the eigendecomposition $\rho =\sum_{l=1}^{d}\lambda
_{l}|\varphi _{l}\rangle \langle \varphi _{l}|$ and a $d^{\prime }\times
d^{\prime }$ unitary matrix $U=(U_{jk})_{jk}$ as \cite{Nielsen-2000-book}
\begin{eqnarray}
\sqrt{p_{k}}|\psi _{k}\rangle =\sum_{l=1}^{d}U_{kl}\sqrt{\lambda _{l}}%
|\varphi _{l}\rangle . \label{eqA.15}
\end{eqnarray}
We can always suppose $d^{\prime }\geq d$ since $p_{k}=0$ is permitted. We
can write $\rho=\sum_{l=1}^{d^{\prime }}\lambda _{l}|\varphi _{l}\rangle
\langle \varphi _{l}|$ by adding $\lambda _{d+1}=\lambda _{d+2}=...=\lambda
_{d^{\prime }}=0$, and $\{|\varphi _{l}\rangle \}_{l=1}^{d^{\prime
}}=\{|\varphi _{l}\rangle \}_{l=1}^{d}\cup \{|\varphi _{l}\rangle
\}_{l=d+1}^{d^{\prime }}$ an orthornormal basis of a $d^{\prime }$%
-dimensional Hilbert space $H^{\prime }.$
Consequently,
\begin{eqnarray}
p_{k} &=&\sum_{l=1}^{d}\lambda _{l}|U_{kl}|^{2},   \label{eqA.16} \\
\langle \psi _{k}|A|\psi _{k}\rangle
&=&\frac{1}{p_{k}}\sum_{l,l^{\prime }=1}^{d}U_{kl}U_{kl^{\prime }}^{\ast }%
\sqrt{\lambda _{l}\lambda _{l^{\prime }}}\langle \varphi _{l}|A|\varphi
_{l^{\prime }}\rangle \nonumber \\
&=&\frac{1}{p_{k}}\text{tr}(Z_{A}\Gamma _{k}), \label{eqA.17}
\end{eqnarray}
with
\begin{eqnarray}
\Gamma _{k}=\sum_{l,l^{\prime }=1}^{d^{\prime }}\sqrt{\frac{\lambda
_{l}+\lambda _{l^{\prime }}}{2}}U_{kl}U_{kl^{\prime }}^{\ast }|\varphi
_{l}\rangle \langle \varphi _{l^{\prime }}|. \label{eqA.18}
\end{eqnarray}

As a result
\begin{eqnarray}
\sum_{k=1}^{d^{\prime }}p_{k}\langle \psi _{k}|A|\psi _{k}\rangle
^{2}=\sum_{k}[\text{tr}(Z_{A}\frac{\Gamma _{k}}{\sqrt{p_{k}}})]^{2}, \label{eqA.19}
\end{eqnarray}
and Eq. (\ref{eqA.14}) becomes
\begin{eqnarray}
\sum_{k=1}^{d^{\prime }}[\text{tr}(Z_{A}\frac{\Gamma _{k}}{\sqrt{p_{k}}}%
)]^{2}\leq \text{tr}(Z_{A}^{2}). \label{eqA.20}
\end{eqnarray}

We know that all Hermitian matrices on $d^{\prime }$-dimensional Hilbert
space $H^{\prime }$ form a $d^{\prime 2}$-dimensional real Hilbert space $%
H_{\dagger }^{\prime }$ with the inner product $\langle A|A^{\prime }\rangle
=$tr$(AA^{\prime })$ for any Hermitian matrices $A,A^{\prime }$ on $%
H_{\dagger }^{\prime }.$ We can check that 
\begin{eqnarray}
&&\text{tr}(\frac{\Gamma _{j}}{\sqrt{p_{j}}}\frac{\Gamma _{k}}{\sqrt{p_{k}}}) \nonumber
\\
&=&\sum_{l,l^{\prime }=1}^{d^{\prime }}\frac{\lambda _{l}+\lambda
_{l^{\prime }}}{2}U_{jl}^{\ast }U_{jl^{\prime }}U_{kl}U_{kl^{\prime }}^{\ast
}=\delta _{jk}, \label{eqA.21}
\end{eqnarray}
this says $\{\frac{\Gamma _{k}}{\sqrt{p_{k}}}\}_{k=1}^{d^{\prime }}$ are
normalized and orthogonal to each other in $H_{\dagger }^{\prime }$, but not a complete orthonormal basis for $H_{\dagger }^{\prime }$ since $H_{\dagger }^{\prime }$ is of dimension $d^{\prime 2}$. Hence equality in Eq. (\ref{eqA.20}) holds if and only if there exist real number $\{r_{k}\}_{k=1}^{d^{\prime }}$ such that
\begin{eqnarray}
Z_{A}=\sum_{k=1}^{d^{\prime }}r_{k}\Gamma _{k}. \label{eqA.22}
\end{eqnarray}

There is a subtlety to be aware of that when for example $p_{1}=0$ in Eq. (\ref{eqA.16}) then  $p_{1}|\psi _{1}\rangle\langle \psi _{1}|=0$ does not appear in the eigendecomposition $\rho=\sum_{k}p_{k}|\psi _{k}\rangle\langle \psi _{k}|$ and Eqs. (\ref{eqA.19},\ref{eqA.20},\ref{eqA.21},\ref{eqA.22}) all exclude $k=1$.

Taking Eqs. (\ref{eqA.11},\ref{eqA.18}) into Eq. (\ref{eqA.22}), we get that for $\{l\}_{l=1}^{d^{\prime }}$ and $%
\{l^{\prime }\}_{l=1}^{d^{\prime }},$
\begin{eqnarray}
2\frac{\sqrt{\lambda _{l}\lambda _{l^{\prime }}}}{\lambda _{l}+\lambda
_{l^{\prime }}}\langle \varphi _{l}|A|\varphi _{l^{\prime }}\rangle
&=&\sum_{k=1}^{d^{\prime }}r_{k}U_{kl}U_{kl^{\prime }}^{\ast }, \label{eqA.23} \\
UY_{A}U^{\dagger } &=&\sum_{k=1}^{d^{\prime }}r_{k}|\varphi _{k}\rangle
\langle \varphi _{k}|, \label{eqA.24}
\end{eqnarray}
where
\begin{eqnarray}
Y_{A}=2\sum_{l,l^{\prime }=1}^{d}\frac{\sqrt{\lambda _{l}\lambda _{l^{\prime
}}}}{\lambda _{l}+\lambda _{l^{\prime }}}|\varphi _{l}\rangle \langle
\varphi _{l}|A|\varphi _{l^{\prime }}\rangle \langle \varphi _{l^{\prime }}|. \label{eqA.25}
\end{eqnarray}

We then conclude that, any $d^{\prime }\times d^{\prime }$ $(d^{\prime }\geq
d)$ unitary matrix $U$ which diagonalizes $Y_{A}$ yields a pure state
ensemble which achieves the equlity in Eq. (\ref{eqA.13}). Apply this result to Eqs. (\ref{eqA.7},\ref{eqA.8}), we
get that $C_{\text{F}}(\rho ,E)=C_{\text{F}}^{C}(\rho ,E)$ if and only if
there exists a $d^{\prime }\times d^{\prime }$ $(d^{\prime }\geq d)$ unitary
matrix $U$ which simultaneously diagonalizes $\{Y_{E_{j}}\}_{j=1}^{n}$. Two
Hermitian matrices $A,A',$ can be simultaneously unitarily diagonalized if and only if
they are commute, i.e., $AA'=A'A.$ Hence, $C_{\text{F}}(\rho ,E)=C_{\text{F}%
}^{C}(\rho ,E)$ if and only if $\{Y_{E_{j}}\}_{j=1}^{n}$ commute to each
other.

For $d=2$ and $\rho=\sum_{l=1}^{2}\lambda _{l}|\varphi _{l}\rangle \langle \varphi
_{l}|$ the eigendecomposition of $\rho,$ from Eq. (\ref{eqA.25}) we have
\begin{eqnarray*}
Y_{|1\rangle \langle 1|} &=&\left(
\begin{array}{cc}
|\langle \varphi _{1}|1\rangle |^{2} & 2\sqrt{\lambda _{1}\lambda _{2}}%
\langle \varphi _{1}|1\rangle \langle 1|\varphi _{2}\rangle \\
2\sqrt{\lambda _{1}\lambda _{2}}\langle \varphi _{2}|1\rangle \langle
1|\varphi _{1}\rangle & |\langle \varphi _{2}|1\rangle |^{2}%
\end{array}%
\right) , \\
Y_{|2\rangle \langle 2|} &=&\left(
\begin{array}{cc}
|\langle \varphi _{1}|2\rangle |^{2} & 2\sqrt{\lambda _{1}\lambda _{2}}%
\langle \varphi _{1}|2\rangle \langle 2|\varphi _{2}\rangle \\
2\sqrt{\lambda _{1}\lambda _{2}}\langle \varphi _{2}|2\rangle \langle
2|\varphi _{1}\rangle & |\langle \varphi _{2}|2\rangle |^{2}%
\end{array}%
\right). \\
\end{eqnarray*}
Notice that $|\langle \varphi _{1}|2\rangle |=|\langle \varphi
_{2}|1\rangle |,|\langle \varphi _{2}|2\rangle |=|\langle \varphi
_{1}|1\rangle |,\langle \varphi _{1}|2\rangle \langle 2|\varphi _{2}\rangle
=-\langle \varphi _{1}|1\rangle \langle 1|\varphi _{2}\rangle ,$ we can
directly check that $\{Y_{|j\rangle \langle j|}\}_{j=1}^{2}$ commute, then $%
C_{\text{F}}(\rho ,\{|j\rangle \}_{j=1}^{2})=C_{\text{F}}^{C}(\rho
,\{|j\rangle \}_{j=1}^{2}).$

For $d=3,$ consider the state
\begin{eqnarray}
\rho &=&\frac{1}{2}|\varphi _{1}\rangle \langle \varphi _{1}|+\frac{1}{2}%
|\varphi _{2}\rangle \langle \varphi _{2}|, \label{eqA.26} \\
|\varphi _{1}\rangle &=&\frac{1}{\sqrt{3}}(|1\rangle +|2\rangle +|3\rangle ), \label{eqA.27}
\\
|\varphi _{2}\rangle &=&\frac{1}{\sqrt{3}}(|1\rangle +e^{i\frac{2\pi }{3}%
}|2\rangle +e^{-i\frac{2\pi }{3}}|3\rangle ). \label{eqA.28}
\end{eqnarray}
From Eq. (\ref{eqA.25}) we have
\begin{eqnarray*}
Y_{|1\rangle \langle 1|} &=&\left(
\begin{array}{ccc}
\frac{1}{3} & \frac{1}{3} & 0 \\
\frac{1}{3} & \frac{1}{3} & 0 \\
0 & 0 & 0%
\end{array}%
\right) , \\
Y_{|2\rangle \langle 2|} &=&\left(
\begin{array}{ccc}
\frac{1}{3} & \frac{1}{3}e^{i\frac{2\pi }{3}} & 0 \\
\frac{1}{3}e^{-i\frac{2\pi }{3}} & \frac{1}{3} & 0 \\
0 & 0 & 0%
\end{array}%
\right), \\
Y_{|3\rangle \langle 3|} &=&\left(
\begin{array}{ccc}
\frac{1}{3} & \frac{1}{3}e^{-i\frac{2\pi }{3}} & 0 \\
\frac{1}{3}e^{i\frac{2\pi }{3}} & \frac{1}{3} & 0 \\
0 & 0 & 0%
\end{array}%
\right).
\end{eqnarray*}
We can directly check that any two of $\{Y_{|j\rangle \langle j|}\}_{j=1}^{3}$ do not
commute, then $C_{\text{F}}(\rho ,\{|j\rangle \}_{j=1}^{3})<C_{%
\text{F}}^{C}(\rho ,\{|j\rangle \}_{j=1}^{3}).$ This example disproves the assertion  $C_{\text{F}}(\rho ,\{|j\rangle \}_{j=1}^{d})=C_{
\text{F}}^{C}(\rho ,\{|j\rangle \}_{j=1}^{d}).$

\bibliographystyle{apsrev4-1}
%

\end{document}